\documentclass[prb,twocolumn,superscriptaddress,showpacs]{revtex4}
\usepackage{amsmath}
\usepackage{amsfonts}
\usepackage{amssymb}
\usepackage{graphicx}
\usepackage[usenames,dvipsnames,svgnames,table]{xcolor}
\usepackage[normalem]{ulem}
\usepackage[makeroom]{cancel}
\usepackage{float}
\usepackage[caption = false]{subfig}
\usepackage[normalem]{ulem}
\usepackage[makeroom]{cancel}
\usepackage{notes2bib}
\usepackage{epstopdf}

\begin{document}

\title{Conductivity of anisotropic inhomogeneous superconductors above critical
temperature.}
\author{S.S.\ Seidov}
\author{K.K. Kesharpu}
\author{P.I.\ Karpov}
\email{karpov.petr@gmail.com}
\affiliation{National University of Science and Technology MISiS, Moscow, Russia}
\author{P.D.\ Grigoriev}
\email{grigorev@itp.ac.ru}
\affiliation{L.D. Landau Institute for Theoretical Physics, 142432, Chernogolovka, Russia}
\affiliation{National University of Science and Technology MISiS, Moscow, Russia}
\affiliation{P.N. Lebedev Physical Institute, RAS, 119991, Moscow, Russia}
\date{\today }

\begin{abstract}
We propose a model and derive analytical expressions for conductivity in
heterogeneous fully anisotropic conductors with ellipsoid superconducting
inclusions. This model and calculations are useful to analyze the observed
temperature dependence of conductivity anisotropy in various anisotropic
superconductors, where superconductivity onset happens inhomogeneously in
the form of isolated superconducting islands. The results are applied to
explain the experimental data on resistivity above the transition
temperature $T_c$ in the high-temperature superconductor $\mathrm{YBa_2Cu_4O_8}$ and in the organic superconductor $\beta$-(BEDT-TTF)$_{2}$I$_{3}$. The comparison of resistivity data and diamagnetic response in $\beta$-(BEDT-TTF)$_{2}$I$_{3}$  allows us to estimate the size of superconducting inclusions as $d\sim 1\mu m$.
\end{abstract}

\pacs{74.81.-g,74.25.fc,74.72.-h,74.70.-b,74.70.Xa,74.70.Kn,74.62.-c}
\maketitle

\section{Introduction}

The appearance of superconductivity with a temperature decrease in many
compounds occurs nonuniformly along the sample. Such an inhomogeneous
superconductivity onset is typical for the majority high-temperature
superconductors (SC), e.g., copper-oxide and iron-based,\cite%
{HighTcNatureReview2015,FeBasedHighTcNatureReview2016,FeSeReview,KresinReview2006}
and it has been directly observed in numerous scanning tunneling microscopy
(STM) experiments on various compounds\cite%
{KresinReview2006,InhBISCCO,InhBISCO2009,InhCaFeAs,InhCaFeAs2018,InhNbN,InhFeSe}%
. The two main reasons for this inhomogeneity are the non-stoichiometry,
coming from doping, and the interplay between different types of electronic
ordering, often leading to phase separation. The diamagnetic response and
the decrease of resistivity far above the superconducting transition
temperature $T_{c}$ are the typical precursors of inhomogeneous
superconductivity,\cite%
{DiaCuprates,DiaPbIn,DiaLaSrCuO,DiaYBCO,DiaYBCOInh,DiaTlBaCuO,Hussey:1997,DiaBaFeCoAs,Grigoriev:2017-PRB,GrigorievJETPL2017,AnisET2I3,PasquierPRB2010,Gerasimenko2014}
which cannot be explained\cite{DiaYBCO,GrigorievJETPL2017} by the standard
theory\cite{LarkinVarlamovFluct} of superconducting fluctuations. Using
scanning SQUID microscopy, such diamagnetic response was even shown to be
highly inhomogeneous and extending far above $T_{c}$.\cite%
{DiaLaSrCuO,DiaYBCOInh} Thus, in 500-nm-thick La$_{2-x}$Sr$_{x}$CuO$_{4}$
films with $T_{c}=18$K the diamagnetic domains of the size $\sim 5-200\mu m$
were observed up to a temperature $80\text{K}\gg T_{c}$ and attributed to isolated
superconducting islands as precursors of superconductivity onset.\cite%
{DiaLaSrCuO} With lowering temperature these superconducting islands become
larger and finally cover most of the area at $T\approx T_{c}$.\cite%
{DiaLaSrCuO} Similar diamagnetic domains of the size $\sim 100\mu m$ above $%
T_{c}$ were also observed in YBa$_{2}$Cu$_{3}$O$_{7-y}$ films.\cite%
{DiaYBCOInh}

The materials, where superconductivity onset shows such spatial
inhomogeneity, are usually characterized by layered crystal structure and
strong anisotropy of electronic properties. The resistivity drop above $T_c$
in these compounds is often much stronger along the least conducting axis%
\cite%
{Hussey:1997,DiaBaFeCoAs,Grigoriev:2017-PRB,GrigorievJETPL2017,AnisET2I3,PasquierPRB2010,Gerasimenko2014},
which again contradicts\cite{GrigorievJETPL2017} (see Appendix \ref%
{app:SCFluctuations} for details) the theory\cite%
{Tinkham,LarkinVarlamovFluct} of superconducting fluctuations in homogeneous
superconductors. This anisotropic effect of incipient superconductivity was
recently explained\cite{Grigoriev:2017-PRB,GrigorievJETPL2017} using a
classical effective-medium model\cite{Torquato} for strongly anisotropic
heterogeneous quasi-2D metal with spheroidal superconducting inclusions,
which is a generalization of the well-known Maxwell's approximation\cite%
{Torquato} for the case of an anisotropic media with non-spherical
inclusions. This simple model predicts\cite%
{Grigoriev:2017-PRB,GrigorievJETPL2017} that if superconductivity in
anisotropic conductors appears in the form of isolated superconducting
islands, it reduces electric resistivity anisotropically with the maximal
effect along the least conducting axis.

The qualitative idea behind this model\cite%
{Grigoriev:2017-PRB,GrigorievJETPL2017} is simple. In a strongly anisotropic
conductor with interlayer conductivity $\sigma_{zz}$ much less than
intralayer conductivity $\sigma_{xx}\sim\sigma_{yy}$, the first,
standard way of interlayer current perpendicular to the conducting layers is
small by the parameter $\eta\equiv\sigma_{zz}/\sigma_{xx}\ll 1$. However,
there is a second way via superconducting islands. Since these islands are
rare, the major part of the current path goes in the normal phase. But
instead of going along the weakly-conducting $z$-axis in the
non-superconducting phase, this second current path between the
superconducting islands goes along the highly conducting layers, until it
comes to another superconducting island, which allows the next lift in the
interlayer direction. Then there is no local current density along the $z$
axis in the non-superconducting phase, and the interlayer conductivity
contribution from this channel does not acquire the small anisotropy factor $%
\sigma_{zz}/\sigma_{xx}\ll 1$. Instead, it acquires another small factor --
the volume fraction $\phi $ of the superconducting phase. If the ratio $\phi
/\eta \gtrsim 1$, the second way makes the main contribution to the
interlayer conductivity.

In Refs. [\onlinecite{Grigoriev:2017-PRB,GrigorievJETPL2017}], analytical
formulas for conductivity in such heterogeneous superconductor were
obtained. These formulas provide a good quantitative agreement with
experimental data on resistivity in FeSe and allow extracting the
temperature dependence of the volume fraction $\phi $ of the superconducting
phase in this compound.\cite{Grigoriev:2017-PRB,GrigorievJETPL2017} If
experimental data on the temperature dependence of diamagnetic response are
available in addition to transport measurements, their comparison also
suggests the approximate shape of superconducting inclusions.\cite%
{GrigorievJETPL2017} However, the obtained expressions\cite%
{Grigoriev:2017-PRB,GrigorievJETPL2017} for conductivity in such a
heterogeneous superconductor are applicable only for the case when
electronic properties in the conducting $a-b$ plane are isotropic. In FeSe
it works well because in spite of the nematic transition at $T\approx 90$K,\cite%
{FeBasedHighTcNatureReview2016,FeSeReview} breaking the $a-b$ isotropy, the
real crystals of FeSe consist of a large number of nanoscale monocrystals
oriented differently along the $a$ or $b$ axis, which restores the $a-b$
isotropy on average. This $a-b$ isotropy in FeSe can be easily broken by
applying a uniaxial pressure.

The limitation to only the isotropic quasi-2D case does not allow application of
the expressions for conductivity of Refs.[%
\onlinecite{Grigoriev:2017-PRB,GrigorievJETPL2017}] to a large number of
superconducting compounds with fully anisotropic electronic properties,
where resistivity along all three main axes differs. Among such fully
anisotropic compounds are most organic metals,\cite{LebedBook,OMRev} where
there are extensive experimental data on resistivity anisotropy above the
superconducting transition temperature $T_c$. These data often show a 	much
stronger effect of incipient superconductivity on interlayer resistivity
above $T_c$,\cite{AnisET2I3,PasquierPRB2010,Gerasimenko2014} which is
qualitatively consistent with the model\cite%
{Grigoriev:2017-PRB,GrigorievJETPL2017} of heterogeneous superconductivity
onset. In many organic superconductors there are quasi-1D Fermi-surface
parts, and superconductivity competes with a charge- or spin-density wave,
leading to their phase coexistence and possible spatial separation in some
pressure interval, as e.g. in (TMTSF)$_{2}$PF$_{6}$,\cite{PasquierPRB2010}
(TMTSF)$_{2}$ClO$_{4}$,\cite{Gerasimenko2014} or $\alpha $-(BEDT-TTF)$_{2}$%
KHg(SCN)$_{4}$.\cite{Andres2005} The type of such phase coexistence and the
corresponding microscopic structure of superconductivity in these compounds
is still debated,\cite%
{Andres2005,PasquierPRB2010,Gerasimenko2014,GGPRB2007,GrigorievPRB2008,GrigPhysicaB2009,GrigPhysicaB2015,Chaikin}
but this density-wave state can be suppressed by pressure of several kbar.%
\cite{LebedBook,OMRev} Some organic superconductors, e.g. $\beta $-(BEDT-TTF)%
$_{2}$I$_{3}$,\cite{AnisET2I3} $\kappa $-(BEDT-TTF)$_{2}$Cu[N(CN)$_{2}$]Br,%
\cite{AnisETBr} $\kappa $-(BEDT-TTF)$_{2}$Cu(NCS)$_{2}$,\cite{AnisNCS} have
only quasi-2D Fermi surfaces, which are anisotropic in the conducting plane
but do not have a nesting property, and hence they are not subject to Peierls or
density-wave instability even at ambient pressure. The temperature
dependence of resistivity anisotropy, shown in Fig. 6 of Ref. [%
\onlinecite{AnisET2I3}], reveals a stronger decrease of interlayer
resistivity $\rho_c$ as compared to $\rho_a$ and $\rho_b$ above the
metal-superconductor transition, which may signify an inhomogeneous
superconductivity onset according to the model of Refs. [%
\onlinecite{Grigoriev:2017-PRB,GrigorievJETPL2017}]. In many cuprate high-$%
T_c$ superconductors, such as YBa$_2$Cu$_4$O$_8$, the chains between
conducting layers break the $a-b$ isotropy. The $a-b$ isotropy in cuprates
may also become broken due to the stripe electronic ordering, as proposed
for La$_{2-x}$Sr$_x$CuO$_4$ ($x=0.02-0.04$) and YBa$_2$Cu$_6$O$_{y}$ ($%
y=6.35-7.0$). \cite{Stripes} In many iron-based high-$T_c$ superconductors
the $a-b$ isotropy is also often broken in the detwinned crystals.\cite%
{AnisFeSe}

In this paper we derive analytical expressions for conductivity in a fully
anisotropic conductors with ellipsoid superconducting inclusions, thus
removing the limitation of in-plane isotropy used in Refs. [%
\onlinecite{Grigoriev:2017-PRB,GrigorievJETPL2017}]. Then we apply our
results to analyze the experimental data on the
temperature dependence of resistivity along three main axes above $T_c$ in the  high-$T_c$ superconductor $\mathrm{YBa_2Cu_4O_8}$ and in the organic superconductor  $\beta$-(BEDT-TTF)$_{2}$I$_{3}$.

The paper is organized as follows. In Sec. II we describe the mapping of the conductivity problem from the anisotropic to the isotropic case. In Sec. III we present our main analytical results for the anisotropic conductivity problem with the superconducting inclusions (some technical details are also described in Appendix B). In Sec. IV we apply the derived analytical results to the analysis of experimental data on $\mathrm{YBa_2Cu_4O_8}$ and $\beta$-(BEDT-TTF)$_{2}$I$_{3}$.
In Sec. V and VI we present a discussion and conclusions.

%

\section{Mapping of conductivity problem in anisotropic media to isotropic}

First, consider a homogeneous anisotropic conducting medium with conductivities  $\sigma^m_{xx}$, $\sigma^m_{yy}$, $\sigma^m_{zz}$ along the principal axes. The electrostatic continuity equation for the
medium can be written as
\begin{equation}
-\mathbf{\nabla j}=\sigma^{m}_{xx}\frac{\partial ^{2}V}{\partial x^{2}}%
+\sigma^{m}_{yy}\frac{\partial ^{2}V}{\partial y^{2}}+\sigma^{m}_{zz}\frac{%
\partial ^{2}V}{\partial z^{2}}=0.
\end{equation}%
Here ${\mathbf j}$ is the current density and $V$ is the electrostatic potential. By the change of the coordinates
\begin{align}
x=x', \quad y=\sqrt{\mu} y', \quad z=\sqrt{\eta}z',  \label{CT}
\end{align}
where
\begin{align}
\mu = \frac{\sigma ^{m}_{yy}}{\sigma ^{m}_{xx}},\quad \eta =\frac{\sigma
^{m}_{zz}}{\sigma^{m}_{xx}}  \label{mu-eta}
\end{align}
and by the simultaneous change of conductivity to $\sigma^{m}=\sigma ^{m}_{xx}$ it transforms to the electrostatic continuity equation
for isotropic media:
\begin{equation}
-\mathbf{\nabla j}=\sigma^{m}\left( \frac{\partial ^{2}V}{%
\partial x'^{2}}+\frac{\partial ^{2}V}{\partial y'^{2}}+\frac{%
\partial ^{2}V}{\partial z'^{2}}\right) =0.
\end{equation}%
Hence, the initial problem of conductivity in anisotropic media with some
boundary conditions can be mapped to the conductivity problem in isotropic
media with new boundary conditions, obtained from the initial ones by the anisotropic dilatation given by Eq. (\ref{CT}).

Second, consider spherical inclusion particles with radii $a_1$ inside the media. Under the transformation (\ref{CT}), these spherical inclusions $%
x^{2}/a_{1}^{2}+y^{2}/a_{1}^{2}+z^{2}/a_{1}^{2}=1$ transform to ellipsoidal ones $%
x^{2}/a_{1}^{2}+y^{2}/a_{2}^{2}+z^{2}/a_{3}^{2}=1$ with semiaxes
\begin{equation}
a_{1},\, a_{2}=a_{1}/\sqrt{\mu }, \, a_{3}=a_{1}/\sqrt{\eta }.
\end{equation}
If $x$ is the direction of highest conductivity of the medium and $z$ is the direction of lowest conductivity (i.e. if $\sigma^m_{xx} > \sigma^m_{yy} > \sigma^m_{zz}$), then $\mu <\eta <1$ and the ellipsoids become $z$-elongated
(i.e. $a_{3}>a_{2}>a_{1}$). Note that generally $\mu $ and $\eta $ can be
temperature-dependent.

If initially the inclusions are not spherical but have an ellipsoidal shape
with the principal semiaxes $a=a_{1}$, $b=\beta a_{1}$ and $c=\gamma a_{1}$, then after
the mapping to the isotropic media these inclusions keep the ellipsoidal shape
but change the principal semiaxes to
\begin{equation}  \label{eq:shape}
a_{1},a_{2}=a_{1}\beta /\sqrt{\mu },a_{3}=a_{1}\gamma /\sqrt{\eta }.
\end{equation}

\section{Conductivity with ellipsoidal superconducting inclusions}

Using the mapping described in the previous section, the conductivity problem of an anisotropic conducive medium with some inclusion particles can be mapped to an effective isotropic media problem with the different shapes of the particles.

Here we consider a medium, e.g. a normal metal, with the isotropic conductivity $\sigma
_{m}$, containing ellipsoidal islands with conductivity $\sigma _{isl}$
and the volume fraction $\phi$. The macroscopic conductivity of the sample $%
\boldsymbol{\sigma ^{\ast }}=diag(\sigma _{xx}^{\ast },\sigma _{yy}^{\ast
},\sigma _{zz}^{\ast })$ in the effective-medium Maxwell's approximation,
applicable for $\phi\ll 1$, can be obtained from (see Eqs.
(18.9) and (18.10) of Ref. [\onlinecite{Torquato}])
\begin{equation}
(1-\phi )(\sigma _{i}^{\ast }-\sigma^{m})+\phi \frac{\sigma _{i}^{\ast
}-\sigma^{isl}}{1+A_{i}(\sigma^{isl}-\sigma^{m})/\sigma^{m}}=0,
\label{eq:cond}
\end{equation}
where $i=1,2,3$ corresponds to $x,y,z$ axes, and coefficients $A_{i}$ are
given by (see Eq. (17.25) of Ref. [\onlinecite{Torquato}]):
\begin{equation}
A_{i}=\frac{a_{1}a_{2}a_{3}}{2}\int\limits_{0}^{\infty }\frac{dt}{%
(t+a_{i}^{2})\sqrt{(t+a_{1}^{2})(t+a_{2}^{2})(t+a_{3}^{2})}}.  \label{El_int}
\end{equation}%
The integrals can be evaluated analytically (see Appendix \ref%
{app:Elliptic_integrals}). For superconducting islands $\sigma
^{isl}\rightarrow \infty $, Eq. (\ref{eq:cond}) simplifies to
\begin{equation}
(1-\phi )(\sigma _{i}^{\ast }-\sigma^{m})-\phi \frac{\sigma^{m}}{A_{i}}%
=0.  \label{eq:cond_lim}
\end{equation}%
Solving it for $\sigma _{i}^{\ast }$ we obtain
\begin{equation}
\sigma _{i}^{\ast }(\phi )=\sigma^{m} \left( 1+\frac{\phi }{A_{i}(1-\phi )}%
\right) \approx \sigma^{m}\left( 1+\frac{\phi }{A_{i}}\right) .
\label{eq:sigma}
\end{equation}
One can also calculate the resistivity
\begin{equation}
\rho _{i}^{\ast }(\phi )=\frac{1}{\sigma _{i}^{\ast }(\phi )}=\frac{1}{%
\sigma^{m}}\frac{A_{i}(1-\phi )}{\phi +A_{i}(1-\phi )}\approx \frac{1}{%
\sigma^{m}}\frac{1}{1+\phi /A_{i}}.  \label{eq:rho}
\end{equation}%
Transforming back from coordinates $(x',y',z')$ of the
isotropic media to the original coordinates $(x,y,z)$ of the anisotropic one
(see Eq. (\ref{CT})), we obtain the final result for resistivities along the three principal axes of
heterogeneous anisotropic media with elliptic superconducting inclusions:
\begin{equation}
\rho _{1}(\phi )= \frac{1}{\sigma^{m}} \frac{A_{1}(1-\phi )}{\phi +A_{1}
(1-\phi )},  \label{eq:rho1}
\end{equation}%
\begin{equation}
\rho _{2}(\phi )= \frac{1}{\mu \sigma^{m}} \frac{A_{2}(1-\phi )}{\phi
+A_{2}(1-\phi )},  \label{eq:rho2}
\end{equation}%
\begin{equation}
\rho _{3}(\phi )=\frac{1}{\eta \sigma^{m}}\frac{A_{3}(1-\phi )}{\phi
+A_{3}(1-\phi )},  \label{eq:rho3}
\end{equation}%
where $A_{i}$ are given by Eqs. (\ref{A1})-(\ref{A3}) for arbitrary ratios
of ellipsoid semiaxes, or by Eqs. (\ref{A1_appr})-(\ref{A3_appr}) for $%
a_{3}\gg a_{1},a_{2}$.

%

\section{Comparison with experiments}

In this section we apply the developed model to the analysis of two superconducting compounds at $T>T_c$, namely, the high-$T_c$ superconductor $\mathrm{YBa_{2}Cu_{4}O_{8}}$ and the organic superconductor $\beta$-(BEDT-TTF)$_{2}$I$_{3}$.

\subsection{High-$T_c$ superconductor $\mathrm{YBa_{2}Cu_{4}O_{8}}$}

Here we analyze the high-$T_c$
superconductor $\mathrm{YBa_2Cu_4O_8}$ at $T>T_c$. In order to do this, we use the
experimental data for the temperature dependence of resistivities $%
\rho_{i}(T)$ along three principal axes, extracted from Fig. 2 of Ref. [%
\onlinecite{Hussey:1997}] (in our notation, axes 1, 2, 3 with the descending
resistivities correspond to axes $b, a, c$ of Ref. [\onlinecite{Hussey:1997}%
]).

In order to calculate conductivity with the correction due to the superconducting
inclusions we, first, need to know the temperature dependencies of conductivity in the
non-superconducting (metallic) phase above $T_{c}$ along the principal axes. These can
be extracted from the experimental data in different ways. For example, if
superconductivity can be suppressed by magnetic field, then conductivity in the high
magnetic field is approximately the same as without field in the metallic
phase. If such data are available, conductivity in the metallic phase
above $T_{c}$ can be extracted as an extrapolation from the high temperatures,
where the effect of superconductivity is absent or negligible. In YBCO the
available magnetic fields are not sufficient to suppress superconductivity,
and we use extrapolation from higher temperatures.

Assuming that at high temperatures the volume fraction of superconducting
inclusions $\phi$ goes to zero sufficiently fast, becoming negligible at $T>250$%
K, we extract the resistivity of the medium from the high-temperature
asymptotic behavior of $\rho_i(T)$: $\rho^m_i (T) \approx \rho_i(T)$ (here $%
\rho_i^m = 1/\sigma_i^m$). Along the $z$ and $y$ axes the resistivity is
approximately linear at high $T$, and we extract $\rho^m_{yy} = (43.5 +
0.772 \text{K}^{-1} \, T) \, \mu\Omega$cm, $\rho^m_{zz} = (6950 + 3.75
\text{K}^{-1} \, T )\, \mu\Omega$cm. For the $x$-axis the total conductivity is
approximately a sum of contributions from conducting planes and chains. The
chain resistivity is not linear but rather a quadratic function of $T$.\cite%
{Hussey:1997} Hence, according to Ref. [\onlinecite{Hussey:1997}], we use
the chain resistivity obtained from $1/\rho_{\text{chain}} = 1/\rho^m_{xx}-1/\rho^m_{yy}$,
with $\rho_{\text{chain}} = (0.5 + 0.00147 \text{K}^{-2} \, T^2) \mu\Omega$%
cm. This gives $\rho^m_{xx} = \rho_{\text{chain}} \rho^m_{yy}/(\rho_{\text{%
chain}} + \rho^m_{yy})$ and $\sigma^m = 1/\rho^m_{xx}$. Using Eq. (\ref%
{mu-eta}), we obtain $\mu(T)$ and $\eta(T)$.

Solving equation (\ref{eq:rho3}) for $\phi$ we get
\begin{equation}
\phi (T)= \frac{1-\eta \rho _{3}\sigma^{m}} {1+\eta \rho_{3}\sigma^{m}(A_{3}^{-1}-1)}.  \label{phi(T)}
\end{equation}
Equations (\ref{eq:rho1}) and (\ref{eq:rho2}) can also be used for the same
purpose of extracting $\phi (T)$, but $\rho _{3}(T)$ has the most pronounced
drop with decreasing $T$ (compared to the linear extrapolation from high to
low temperatures), so it should give the most accurate results. The
least-conductive $z$-direction has the smallest coefficient among $A_i$ (see
Eqs. (\ref{A1_appr2})-(\ref{A3_appr2})). Hence, according to Eq. (\ref{eq:sigma}%
), $\sigma_{zz}$ is the most sensitive to the concentration of inclusions $%
\phi$, unless the shape of superconducting inclusions is too compressed
along the $z$-axis. Conductivity $\sigma_{xx}$ along the highest conductive
direction is expected to be the least sensitive to the variation of $\phi$.

\begin{figure}[tbh]
\centering
\includegraphics[width=0.9\linewidth]{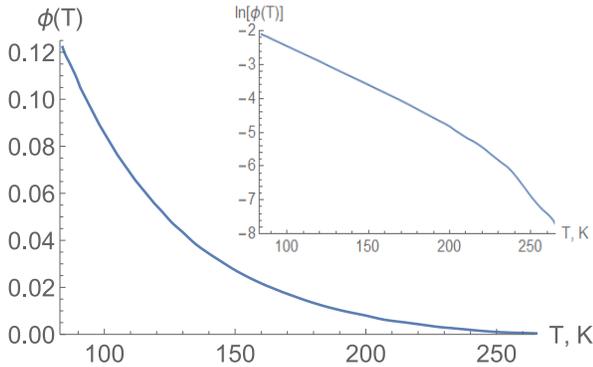}
\caption{Temperature dependence $\phi (T)$ of the volume fraction of
superconducting inclusions extracted from formula (\ref{phi(T)}) and
the experimental data from Ref. [\onlinecite{Hussey:1997}] for $\mathrm{YBa_{2}Cu_{4}O_{8}}$.}
\label{fig_phi(T)}
\end{figure}

\begin{figure}[tbh]
\centering
\subfloat[(a)]{\includegraphics[width=0.95\linewidth]{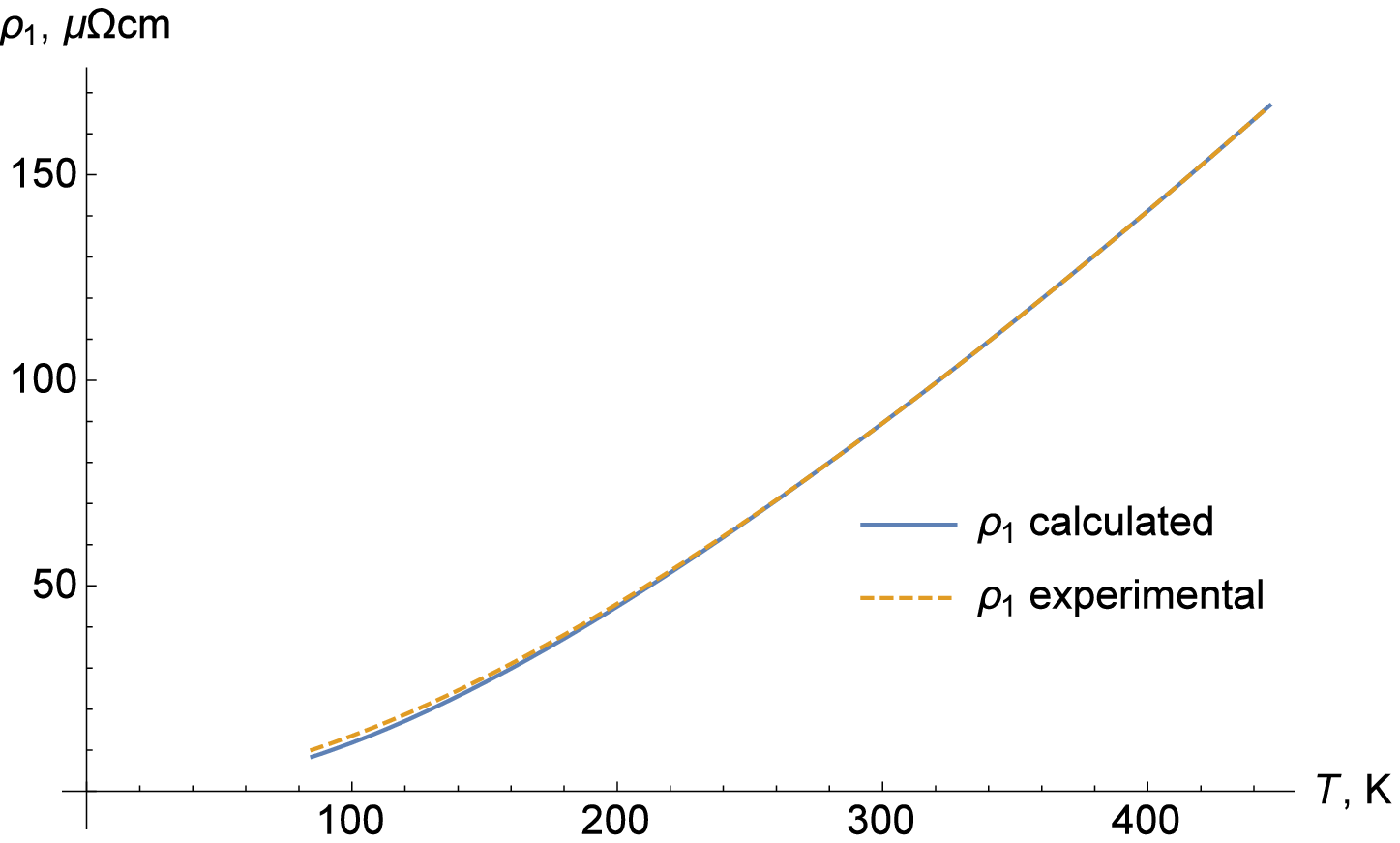}}
\par
\subfloat[(b)]{\includegraphics[width=0.95\linewidth]{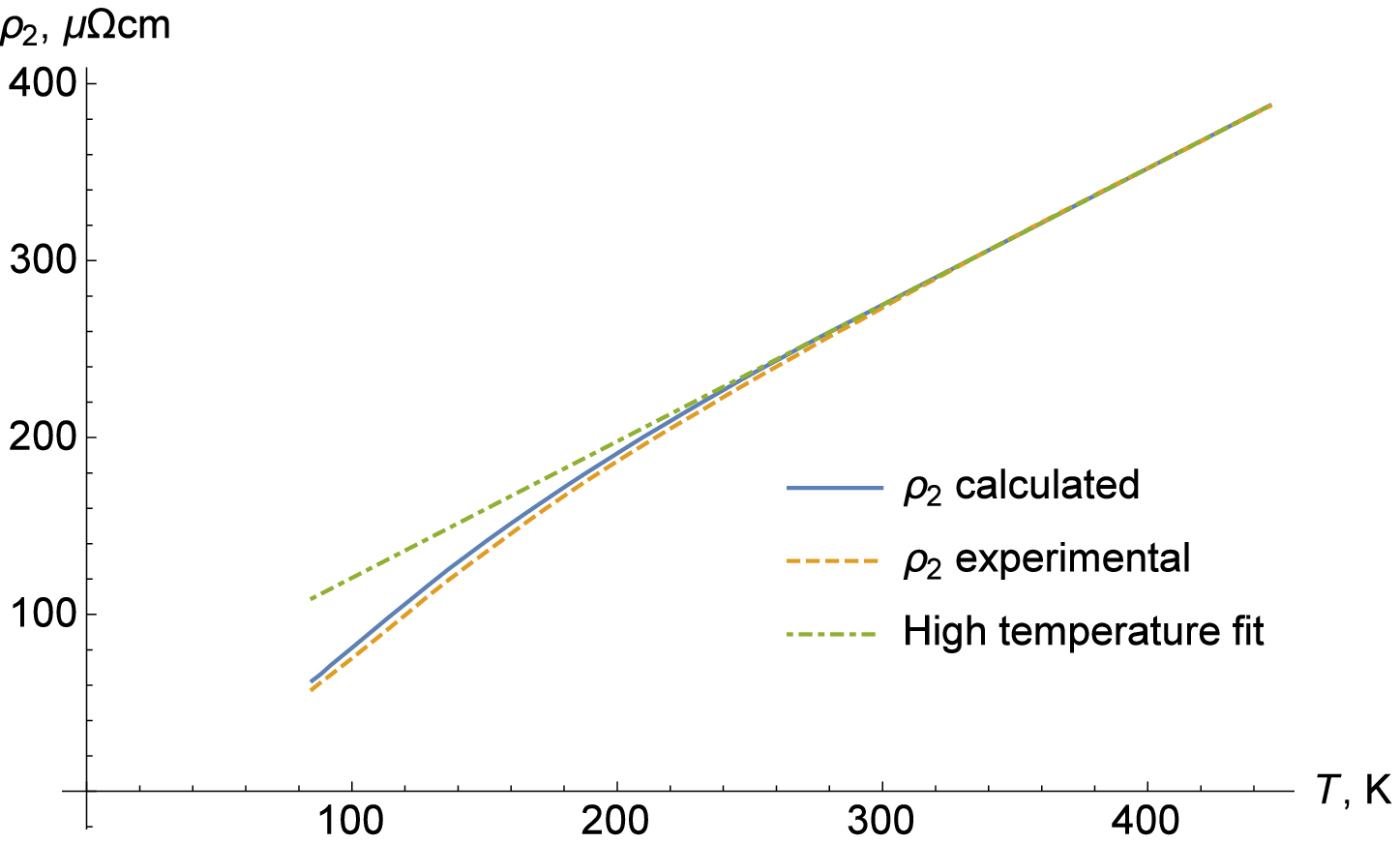}}
\par
\subfloat[(c)]{\includegraphics[width=0.95\linewidth]{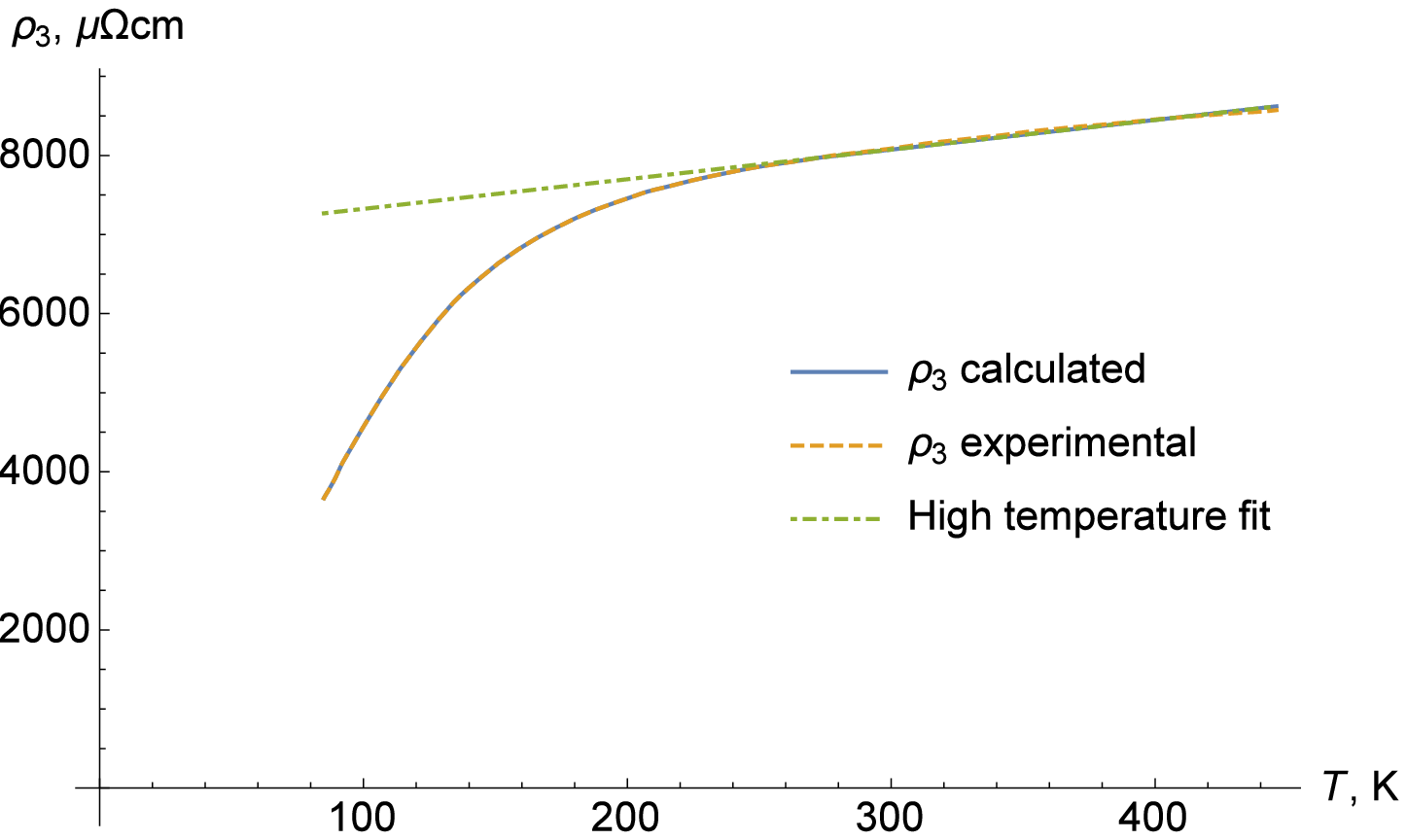}}
\caption{Comparison between the proposed theory and the experimental data from
Ref. [\onlinecite{Hussey:1997}] on the temperature dependence of resistivity in $\mathrm{YBa_{2}Cu_{4}O_{8}}$
along three main axes, $\protect\rho _{i}(T)$ for $i=1,2,3$. In Fig. (a) the high-temperature fit coincides with experimental data and is not shown.}
\label{fig_rho(T)}
\end{figure}

Assuming that the spatial extensions of the superconducting inclusions are
proportional to the coherence lengths $\xi_i \sim v^F_i$ along the
corresponding axes, we take the anisotropy parameters $\beta = v^F_y/v^F_x
=1 $, because superconductivity comes from the conducting planes (not
chains) where the electron dispersion is isotropic, and $\gamma \approx
v^F_z/v^F_x \approx\sqrt{\sigma_{zz}/\sigma_{xx}} \approx 0.15$.
This aspect ratio $\gamma $ is very close to the aspect ratio
$\gamma_o \approx 0.14$ giving the best fit of the resistivity curves within our model.

Figure \ref{fig_phi(T)} shows the temperature dependence $\phi (T)$, derived
from Eq. (\ref{phi(T)}) for the inclusions anisotropy parameters $\beta = 1$
and $\gamma = 0.15$. As expected, $\phi (T)$ decreases with the increase of
temperature and becomes negligibly small (or zero) only at $T\gtrsim 250$K,
which is much higher than the transition temperature $T_{c}=78$K. As one can
see from the inset in Fig. \ref{fig_phi(T)}, where $\phi (T)$ is plotted in
the logarithmic scale, the dependence $\phi (T)$ is nearly exponential in
the temperature range $80$K$<T<200$K. This exponential decrease of $\phi (T)$
with increasing $T$ is natural for the model of isolated superconducting
inclusions coming from disorder or electronic phase separation, but it
contradicts the prediction from the theory
of superconducting fluctuations in homogeneous superconductors \cite{Tinkham,LarkinVarlamovFluct}.

From $\phi (T)$ we derive $\rho _{i}(T)$ using formulas (\ref{eq:rho1})-(\ref%
{eq:rho3}) and in Fig. \ref{fig_rho(T)} we compare the obtained dependencies
with the experimental data of Ref. [\onlinecite{Hussey:1997}]. The
calculated dependence $\rho_3(T)$ trivially coincides with the experimental
one, since we extracted $\phi(T)$ from $\rho_3$. Also, naturally we get a
good agreement for $\rho_1(T)$, since resistivity in the highest conductivity $x$%
-direction depends only weakly on $\phi$, as explained above. The most
important role here is played by $\rho_2(T)$-dependence: the difference
between the high-temperature fit (green dot-dashed line) and our theoretical
prediction (blue solid line) comes from isolated superconducting inclusions
according to the the proposed model, which fits very well with the
experimental data (orange dashed line).

\subsection{Organic superconductor $\beta$-(BEDT-TTF)$_{2}$I$_{3}$}

In this subsection, we apply our theoretical model to analyze the
observed temperature dependence of conductivity anisotropy in a quasi-2D
organic charge transfer salt $\beta$-(BEDT-TTF)$_{2}$I$_{3}$ with superconducting  transition temperature $T_{c}\approx 1.5K$\cite%
{AnisET2I3,Yaguskii1984A,Kaminskii1983}. This compound is convenient for the analysis  because
(i) both resistivity along all three main axes and susceptibility data are
available for it, and (ii) it does not have several complicating
features characteristic of high-$T_c$ cuprate superconductors.

\begin{figure}[tbp]
\includegraphics[width=1.0\linewidth]{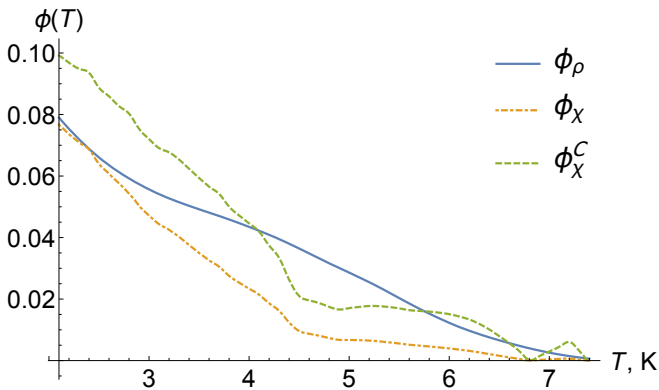}
\caption{Comparison of the temperature dependence of SC volume fraction $\phi$ in $\beta$-(BEDT-TTF)$_{2}$I$_{3}$ calculated from the resistivity data\cite{AnisET2I3} ($\phi_{\rho}$, solid blue curve, Eq. (\ref{phi(T)})), and from the magnetic susceptibility data\cite{Merzhanov1985} using Eq. (\ref{chi1}): with temperature-independent constant $C=2600$ ($\phi_{\chi}$, orange dot-dashed curve) and with temperature-dependent coefficient $C(T)$ given by Eq. (\ref{C}) ($\phi_{\chi}^C$, green dashed curve). }
\label{plot_phi_3}
\end{figure}

We have taken the resistivity data $\rho_{i}(T)$ along three principal axes, extracted from Figs. 2, 3, 4 of Ref. [%
\onlinecite{AnisET2I3}] (in our notation, axes 1, 2, 3 with the descending
resistivities correspond to axes $a, b, c$ of Ref. [\onlinecite{AnisET2I3}]).
The data from the sample denoted by the open circles ($\circ$) were used
\cite{footnote_open_circles}.

Unfortunately, we have not found experimental data on resistivity in this
compound under very high magnetic field, which prevents SC island formation.
Therefore, to find the temperature dependence of metallic conductivity, we
extrapolated resistivity along the highest conductivity $x$-axis, from high temperatures $T > 9$K down to $T=2\div9$K as: $\rho^m_{xx}\equiv\rho^m_{1}=(1.429+0.084\text{K}^{-1} \,T+0.006\text{K}^{-2} \,T^2)\rho_1(293\text{K})\times 10^{-3}$ (where the value of $\rho_1(293\text{K})\approx 54.15$ m$\Omega$cm was extracted from Ref. [\onlinecite{AnisET2I3}]).
In this extrapolation we keep both linear and quadratic terms, which may come from the electron-electron interaction at low temperature. The volume fraction $\phi (T)$ of SC islands is found
from the resistivity $\rho _{3}$ along the lowest conductivity $z$-axis, using Eq. (\ref{phi(T)}) for the SC inclusions of the ellipsoidal shape with the principal semiaxes ratios $a_2=a_3=3a_1$ in the mapped space, or $\beta =b/a \approx 2$, $\gamma = c/a \approx 0.13$ in real coordinate space. The parameters $\beta$ and $\gamma$ were found by minimizing the difference between the theoretical prediction and
the experimental data for resistivity along the $x$ and $y$ axes.
The result for $\phi (T)$ is shown in Fig. \ref{plot_phi_3} by solid blue curve.

\begin{figure}[htb]
\centering
\subfloat[(a)]{\includegraphics[width=0.95\linewidth]{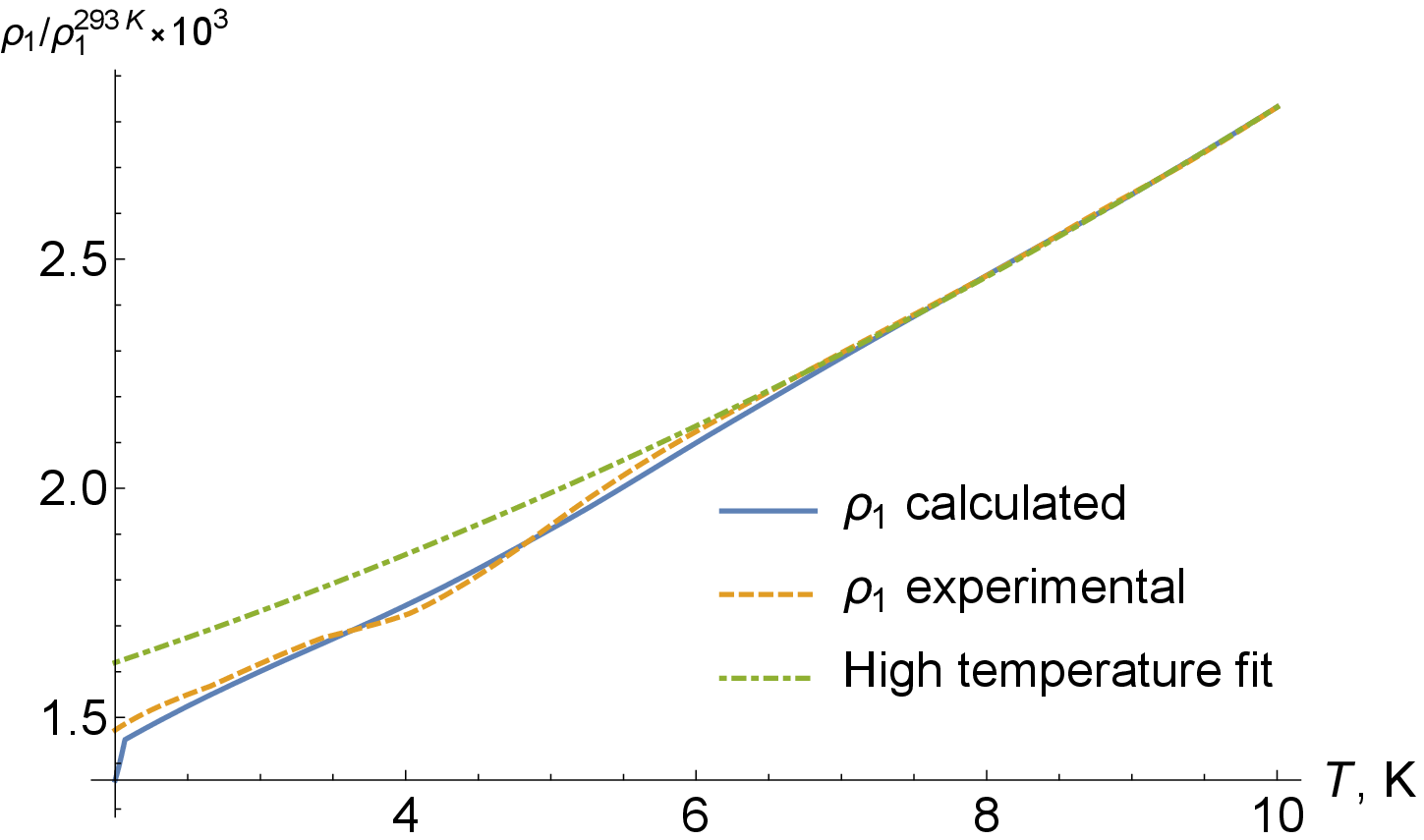}}
\par
\subfloat[(b)]{\includegraphics[width=0.95\linewidth]{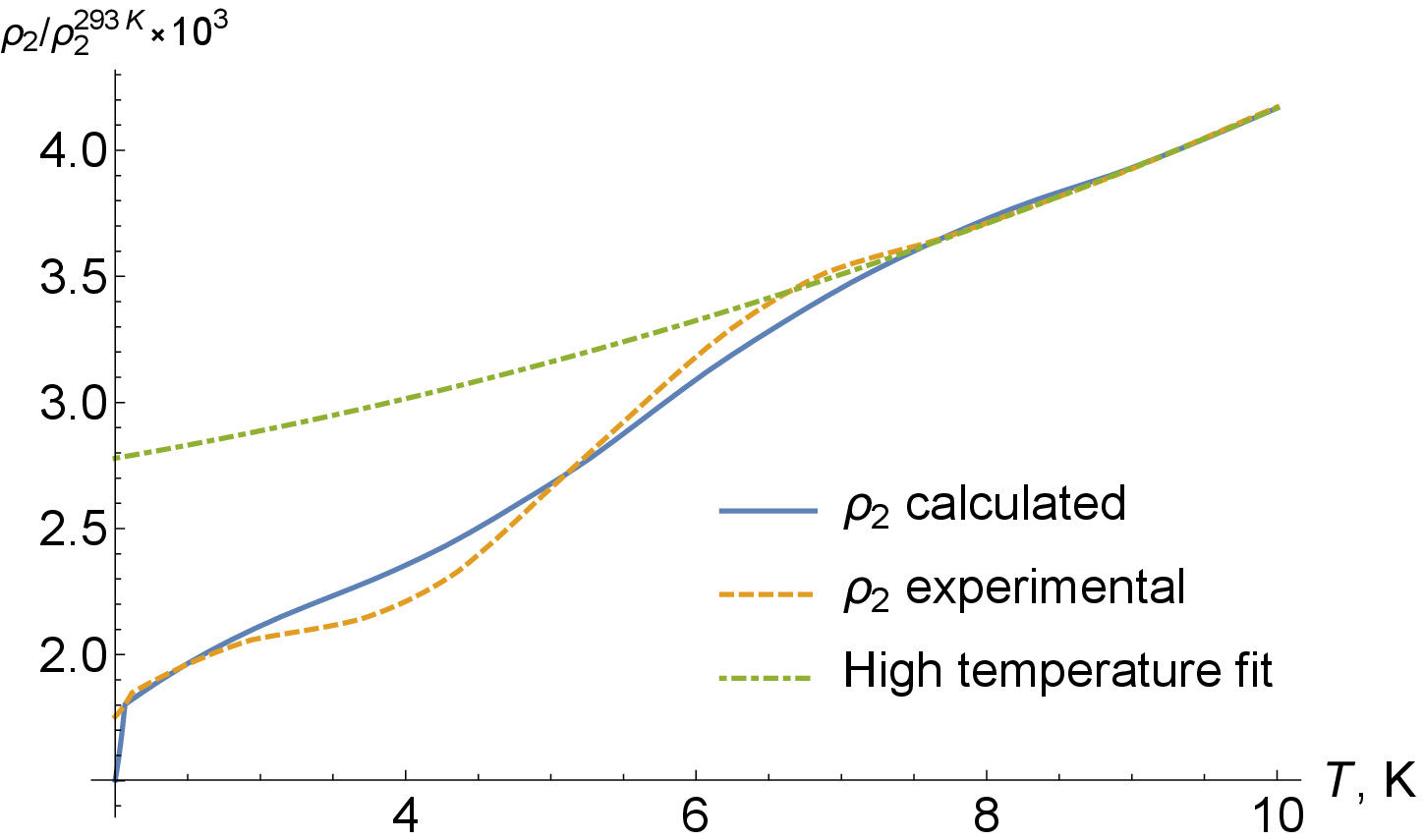}}
\par
\subfloat[(c)]{\includegraphics[width=0.95\linewidth]{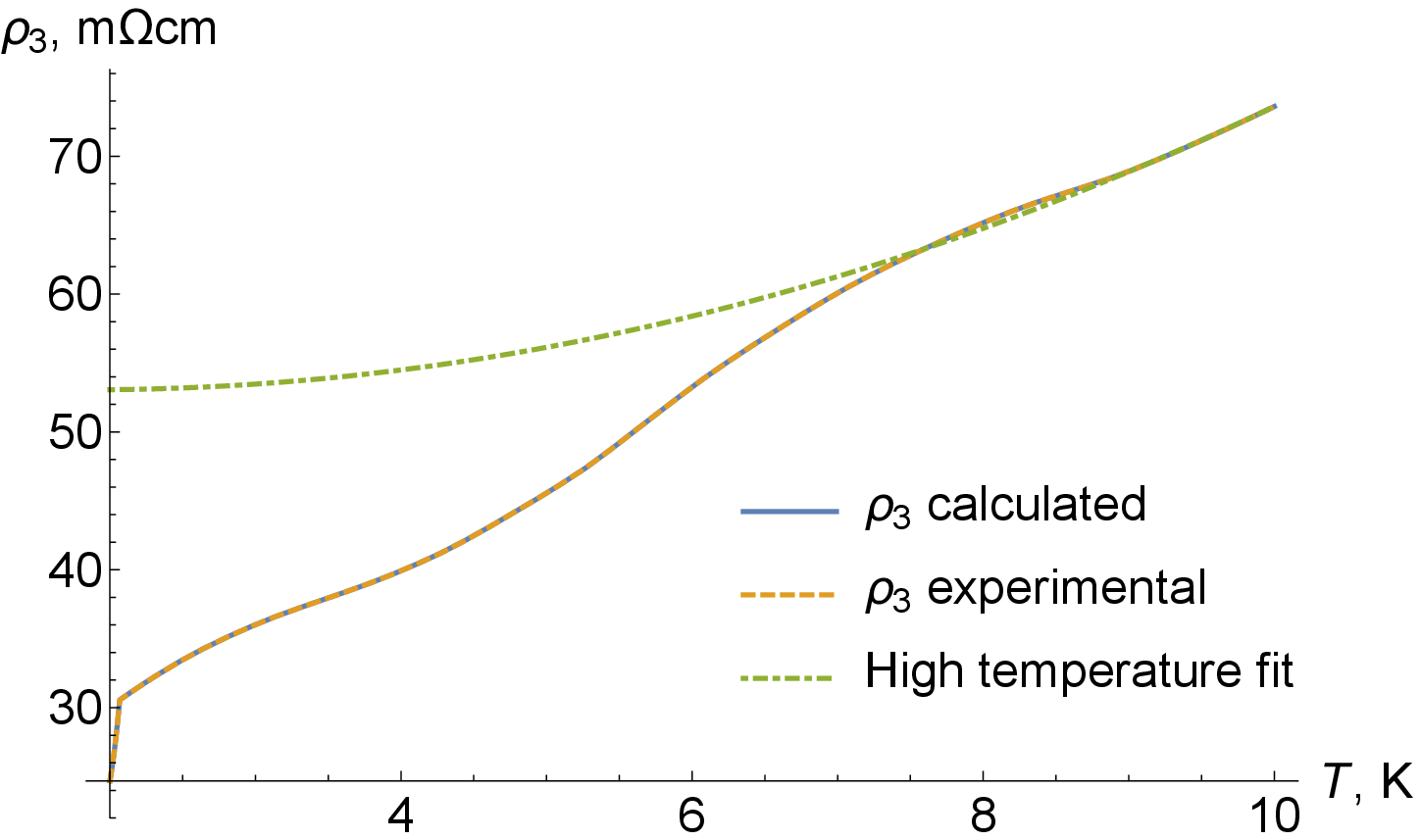}}
\caption{Comparison between the proposed theory and the experimental data from
Ref. [\onlinecite{AnisET2I3}] on the temperature dependence of resistivity in $\beta$-(BEDT-TTF)$_{2}$I$_{3}$
along three main axes, $\rho _{i}(T)$ for $i=1,2,3$.
The high-temperature fit is the resistivity extrapolated from $T>9$K down to low temperatures and given by the second-order polynomials in $T$: $\rho^m_1(T) =(1.429+0.084\text{K}^{-1} \,T+0.006\text{K}^{-2} \,T^2)\rho_{1}(293\text{K})\times10^{-3}$, $\rho^m_2(T) = (2.616+0.063\text{K}^{-1} \,T+0.009\text{K}^{-2} \,T^2)\rho_{2}(293\text{K})\times 10^{-3}$, and $\rho^m_3(T) = (47.821+0.025\text{K}^{-1} \,T+0.255\text{K}^{-2} \,T^2)$m$\Omega$cm.}
\label{plot_resistivity_phi_3}
\end{figure}

In Fig. \ref{plot_resistivity_phi_3} we compare the experimental data on resistivity with the predictions of our model. The experimental and theoretical curves for $\rho _{3}(T)$ trivially coincide because we used $\rho_3(T)$ data to obtain $\phi (T)$ using  Eq. (\ref{phi(T)}). The calculated temperature dependence of two other resistivity components $\rho _{1}(T)$ and $\rho _{2}(T)$, given by solid blue curves in Fig. \ref{plot_resistivity_phi_3}, agrees well with the experimental data (dashed orange curves). The values of resistivity used in Fig. \ref{plot_resistivity_phi_3}(a) and \ref{plot_resistivity_phi_3}(b) at $T=293$K along $a$-axis ($\rho_1^{293 \text{K}} = 54.15m\Omega$cm) and along $b$-axis ($\rho_2^{293 \text{K}}=86.64m\Omega$cm) respectively are calculated from the experimental data given in Ref. [\onlinecite{AnisET2I3}]. The high-temperature fit without SC inclusions is given by the green dot-dashed curve.

We can compare, at least qualitatively, the temperature dependence of the SC
volume fraction $\phi _{\rho }(T)$, calculated from the resistivity data, with
SC volume fraction $\phi _{\chi }(T)$, calculated from the magnetic
susceptibility data, taken from Fig. 1 of Ref. [\onlinecite{Merzhanov1985}]. In this figure the
magnetic susceptibility $\chi _{2.8}$ in a weak magnetic field of $2.8$ kOe has a pronounced drop. This drop starts at $T^*\approx 8$K and was ascribed to incipient superconductivity, because in higher magnetic field $8.0$ kOe the observed diamagnetic susceptibility $\chi _{8.0}$ does not have such a drop.\cite{Merzhanov1985} The standard theory of SC fluctuations\cite{LarkinVarlamovFluct} does not explain such a large difference between $T^*\approx 8$K and the SC transition temperature $T_c\approx 1.5$K. Hence, also taking into account the resistivity data\cite{AnisET2I3}, we suppose that the observed diamagnetic response originates from the SC islands, which survive only at lower magnetic field. Assuming that the low magnetic field $2.8$ kOe almost does not affect the SC inclusions, and in high magnetic field $8$ kOe, on the contrary, the effect of SC islands is negligible, we find that $\phi _{\chi }$ is proportional to the difference $\Delta \chi \equiv (\chi _{{2.8}}-\chi _{{8}})$, divided by
the susceptibility $\chi _{SC}=-1/4\pi $ of a perfect superconductor:
\begin{equation}
\phi _{\chi }=C\Delta \chi /\chi _{SC},  \label{chi1}
\end{equation}%
In Fig. \ref{plot_phi_3} we compare SC volume fraction $\phi _{\rho }(T)$ (blue solid curve) extracted from the resistivity data \cite{AnisET2I3} with $\phi _{\chi }(T)$ (orange dot-dashed curve), determined according to Eq. (\ref{chi1}) with the constant coefficient $C=2.6 \times 10^{3}$, found assuming that around 1.8-2K the values of $\phi_{\rho}(T)$ and $\phi_{\chi}(T)$ on average become close to each other.

The coefficient $C$ is not equal to unity mainly because of three effects: (i) the demagnetizing
factor $n$ of the SC ellipsoids, (ii) the finite penetration depth $\lambda $ of a magnetic field into the SC granules, and (iii) the penetration of vortices if the applied magnetic field exceeds the lower critical field $H_{c1}$.
Susceptibility of a macroscopic SC
ellipsoid is given by $\chi _{ellipsoid}=(-1/[4\pi (1-n)])$,
\cite{Cape1967,Landau1984}.
The demagnetization factor $n$ along the longest semi-axis of the ellipsoid with $b/a = 2$ and $c/a=0.13$ is $n\approx 0.05$, as calculated from the Eq. (2.1) of Ref. [\onlinecite{ellipsoid}]. \cite{footnote}
Hence, the first effect only gives a factor $(1-n)^{-1} \approx 1.05 \sim 1$.

The second effect is more important if the penetration depth $\lambda $ is comparable to or greater than the size $d=2R$ of
SC islands. Using the well-known expression for the diamagnetic
susceptibility $\chi =-R^{2}/40\pi \lambda ^{2}$ of a small spherical SC
granule with radius $R\ll \lambda $ (see Eq. (8.22) of Ref. [\onlinecite{Tinkham}]), we can approximate the measured susceptibility decrease $\Delta \chi $ by
\begin{equation}
\Delta \chi \approx \frac{\chi _{SC}\phi _{\chi }R^{2}}{10\lambda ^{2}(1-n)}.
\label{chi2}
\end{equation}
The penetration depth $\lambda $ depends on $T$ and diverges at the critical temperature
(see Eq. (2.3) of Ref.[\onlinecite{Tinkham}]): $\lambda \left(
T\right) \approx \lambda \left( 0\right) /\sqrt{1-\left( T/T_{c}\right) ^{4}}
$. Here, instead of the macroscopic  zero-resistance value $T_c$ we use the temperature $T^*$ below which SC islands start to appear. This gives the temperature dependence of the constant $C$ in Eq. (\ref%
{chi1}) at $R\ll \lambda $:
\begin{equation}
C\left( T\right) \approx \frac{10\lambda ^{2}(1-n)}{R^{2}}\approx \frac{10%
        \left[ \lambda \left( 0\right) \right] ^{2}(1-n)}{R^{2}\left[ 1-\left(
        T/T^{*}\right) ^{4}\right] }.  \label{C}
\end{equation}
For our estimate of the SC islands radius $R$ we take $T^*\approx 8$K because
(i) there is a clear drop of interlayer resistivity at $T = 6-8$K, suggesting
the appearance of many SC islands below $T^*\approx 8$K, and (ii) this $T^*$ is close to the
SC transition temperature $T_c\approx 7.5$K of another group of crystals\cite{AnisET2I3}
of the same compound but prepared in a different way. Moreover, this simplification only leads to
a minor error in our estimate of the size of SC islands given below, because for this estimate
we compared the SC volume fraction at $T=2-4$K (see Fig. 3), where the factor $[1-(T/T^*)^4]\sim 1$
 is not very sensitive to the precise value of $T^*$.

Eq. (\ref{C}) explains why
at high temperatures $T>4K$, when $\lambda \left(
T\right) /\lambda \left( 0\right) \gg 1$, there is a strong difference
between $\phi _{\rho }$ and $\phi _{\chi }$ calculated for a constant $C$.
Therefore it is more physically motivated to use the SC volume fraction $\phi_{\chi}^{C}(T)$ which takes into account the temperature dependence of the coefficient $C(T)$.
We extract $\phi_{\chi}^{C}(T)$ from the
experimental data \cite{Merzhanov1985} on susceptibility according to Eqs.
(\ref{chi1}) and (\ref{C}) with $\lambda(0)/R=16$; it agrees with
$\phi _{\rho }(T)$ much better than for the temperature-independent coefficient $C$, as shown in Fig. \ref{plot_phi_3}.
Thus we can estimate the typical size of the SC islands as $d=2R\approx \lambda(0)/8$. The usual in-plane London penetration depth in organic metals is rather large, $\lambda(0)\gtrsim 1\mu m$, while the out-of-plane penetration depth is even  $\sim 30$ times larger.\cite{lambdaOM} Using the in-plane London penetration depth $\lambda(0)\approx 6\mu m$ of the compound $\alpha$-(BEDT-TTF)$_{2}$I$_{3}$ from the same family,\cite{lambdaOM} we obtain the typical size $d\approx 0.75\mu m \sim 1 \mu m$ of SC inclusions in $\beta$-(BEDT-TTF)$_{2}$I$_{3}$. This is much greater than the in-plane SC coherence length $\xi_{\parallel}\approx 10-80nm$ \cite{Bulaevskii1987} in this compound. Thus the proximity effect and Josephson coupling give only small corrections to our formulas.

There is a third effect, influencing the diamagnetic response and the the coefficient $C$ in Eq. (\ref{C}). Due to the penetration of magnetic vortices, if the applied magnetic field exceeds the lower critical field $H_{c1}$, then the diamagnetic signal of a bulk superconductor is much smaller than that of an ideal diamagnet. However, if the size of SC islands $R$ is small, much smaller than SC penetration depth $\lambda$, the penetration of magnetic vortices to SC may be energetically unfavorable, so that no or very few vortices are in the small superconducting islands. The effective lower critical field in a thin cylinder of radius $R$ at $\xi\ll R\ll \lambda $ was shown to increase $H_{c1}\propto (\lambda/R)^2$ (see Eq. (4) of Ref. [\onlinecite{Shapoval1999}]) and even exceed upper critical fields $H_{c2}$ and $H_{c3}$ at $R<1.5\xi $ (see Fig. 1 of Ref. [\onlinecite{Shapoval1999}]). For $\lambda /R\approx 16$, as we estimated above, the magnetic field must exceed the bulk $H_{c1}$ more than 300 times for the penetration of a single vortex becoming energetically favorable, which is, probably, not the case in the experiment in Ref. [\onlinecite{Merzhanov1985}], although the applied magnetic field there is much larger than $H_{c1}$ in a bulk superconductor.
If the applied magnetic field in Ref. [\onlinecite{Merzhanov1985}] exceeds this enlarged $H_{c1}$, a few vortices may penetrate the SC islands and reduce the diamagnetic response. Then our estimate of the SC island radius $R$ from Eqs. (\ref{chi2}) and (\ref{C}) gives a lower bound for $R$. This is also helpful, because this lower bound is still larger than the SC coherence length, thus substantiating the applicability of our model.

\section{Discussion}

In order to calculate the classical conductivity of heterogeneous media we have
used the Maxwell's or Maxwell-Garnett approximation, generalized for
anisotropic media. It is valid only in the limit of low volume fraction $%
\phi $ of the second phase, i.e. of superconducting inclusions in our case.
In particular, it gives an incorrect percolation threshold $\phi =1$. However,
Maxwell's approximation has several important advantages: (i) it is
exact in the limit $\phi \ll 1$; (ii) it coincides with the optimal
Hashin-Shtrikman bounds,\cite{Sahimi} i.e., it equals to the lower bound for
the effective conductivity of media with superconducting inclusions for
arbitrary $\phi $; (iii) it does not require the usually unknown information
about the distribution function of superconducting islands and about their
typical size $d$; (iv) it gives a simple analytical result.

There are several other approaches to this classical conductivity problem that
have their own advantages and drawbacks.\cite{Torquato,Sahimi} Among the most
popular analytic approaches, we have the
self-consistent effective-medium approximation, the cluster expansions and
the contrast expansions, giving various bounds for the effective
conductivity tensor. The self-consistent effective-medium approach is the
simplest one after the Maxwell's approximation. It gives nontrivial
percolation thresholds in 2D and 3D cases, which are close to the numerical
results for the isotropic case. However, for anisotropic systems these
percolation thresholds differ for different directions, which is incorrect
in a general case. The self-consistent approximation does not have a strict
substantiation even in the low-$\phi $ limit, but it was shown to describe
correctly some fractal inhomogeneous structures, which are similar on
different length scales.\cite{Torquato,Sahimi} Thus, it is not clear if the
self-consistent or Maxwell's approximation gives better accuracy for our
strongly anisotropic case.

The cluster expansions coincide with Maxwell's approximation for the
dilute dispersions of superconducting islands or in the first order in $\phi $ and give
better accuracy in the higher orders. However, this approach requires the
distribution and correlation functions of superconducting inclusions, which
are unknown. In addition, the cluster expansions do not usually give simple
analytical formulas. The contrast expansions work well when the
conductivities of two phases do not differ much, which is not applicable to
our case where the conductivity ratio is infinite. In addition, the contrast
expansions also require the knowledge of the correlation function and do not
give simple analytical results. The various numerical methods of calculating the
effective conductivity of such a heterogeneous classical system\cite{Sahimi}
give much more accurate results but are not convenient for the physical
analysis. In addition, the accuracy of our classical model is limited,
especially without knowledge of size the distribution and the correlation
function of superconducting islands. Thus, among various methods, the
applied Maxwell-Garnett approximation seems to be reasonable for our
qualitative study.

The applied classical model does not take into account several quantum
superconducting features: the Andreev reflection, the proximity effect, and
the Josephson coupling between superconducting islands.\cite{Tinkham} All
these three effects increase the electric conductivity in such a heterogeneous
medium

The Andreev reflection increases an electric current through the
normal-superconductor (N-S) interface. The increase depends on the
strength of the potential barrier on this interface (see Sec. 11.5.1 of Ref. [\onlinecite%
{Tinkham}]), but it does not exceed the factor of 2. Since in our case the superconducting
islands are made of the same material as the metallic matrix, the potential
barrier at their interface is not large, and the Andreev reflection may almost
double the current through a flat N-S interface. For an ellipsoidal or
arbitrary shape of SC granules this increase factor is less than 2 and
closer to unity; it should be taken into account in a rigorous quantitative
theory, but is beyond our study.

The proximity effect creates a non-zero superconducting condensate in the
surrounding shell of the thickness $\sim \xi $, around each of the superconducting
islands. Since $\xi \propto v_{F}$ has the same anisotropy as electron
velocity $v_{F}$, in layered compounds this shell is thicker along the
conducting layers and thinner along the interlayer $z$ direction. When the
typical size $d$ and distance $l$ between SC islands are greater than $\xi $%
, the proximity effect is qualitatively equivalent to the effective increase
of the SC island size by a length $\sim \xi $. Then it increases the
calculated correction $\Delta \sigma _{i}$ to conductivity along the axis $i$
due to SC islands by a quantity $\sim \left( \xi _{i}/d_{i}\right) \Delta
\sigma _{i}$. For small SC islands of the size $d\lesssim \xi $ the
proximity-effect correction to $\Delta \sigma _{i}$ is not small and must be
taken into account in a quantitative theory, because it increases the
effective volume fraction $\phi $ of SC phase and changes the effective
shape of SC islands, making them closer to an ellipsoid with the main axes $%
d_{i}^{\ast }\propto \xi _{i}$. For naturally inhomogeneous superconductors,
when both SC and metallic regions consist of the same compound, the size of
SC islands $d\gtrsim \xi $, and our analysis remains valid. For example, in $\beta $-(BEDT-TTF)$_{2}$I$_{3}$ the typical size of SC inclusions $d \sim \lambda/8 (0) \sim 1\mu m\gg \xi_{\parallel}\sim 10-80nm \gg \xi_{\perp}\sim 1nm$, and the proximity effect gives only a small correction. However, even in
the case $d_{||}\lesssim \xi_{||} $ the qualitative effect that the strongest
relative increase of conductivity due to SC islands is along the least
conducting axes of metallic matrix may persist if $d_{\perp}\gtrsim \xi_{\perp}$.

For the small inter-island distance $l\lesssim \xi $, the Josephson coupling
between superconducting inclusions becomes important. It gives the phase coherence
to the SC condensates on the neighboring islands and may even lead to
superconductivity of the whole sample if this phase coherence is long-range.
The conductivity of an array of SC granules in a dielectric medium has been
extensively studied in various regimes and the corresponding
superconductor-insulator phase diagram has been obtained theoretically and
experimentally (see Ref. [\onlinecite{BeloborodovRMP}] for a review). Arrays of SC
granules in a metallic matrix received less attention but have also been
investigated in artificial\cite{SternfeldPRB2005} and natural\cite%
{PureurPRB1993,Kresin2003,Ponta} systems. The metal-superconductor
transition in these systems occurs in two stages. First, with lowering
temperature, at $T<T_{c}^{\ast }$, superconductivity appears in isolated
granules, which reduces electric resistivity and gives a diamagnetic response.
At lower temperature, the long-range coherence
between isolated SC islands or clusters are established, and at the resistive
transition temperature $T_{c}<T_{c}^{\ast }$ the whole sample becomes
superconducting. For a random spatial and $T_{c}^{\ast }$ distribution of SC
islands ,this leads to a continuous decrease of resistivity between $%
T_{c}^{\ast }$ and $T_{c}$. Finite-temperature effects break this coherence
when $T$ becomes comparable to the Josephson coupling energy $E_{J}\equiv
\hbar I_{c}/2e$, where $I_{c}$ is the critical current of the Josephson
junction,\cite{Tinkham} which depends exponentially on the intergranular
distance $l$: $I_{c}\propto \exp \left( -l/\xi \right) $. Near $T_{c}^{\ast
} $ of SC granules the critical current has a linear temperature dependence,
$I_{c}\propto T_{c}^{\ast }-T$, and for $l>\xi $ it acquires additional
exponential temperature damping. Hence, in our limit of low volume ratio $%
\phi \ll 1$ of the SC phase, where $l>\xi $, the Josephson coupling is most
probably suppressed by temperature and can be neglected.

Thus, the applied model is quantitatively valid only in the macroscopic limit,
when the size of superconducting islands $d$ and the distance between them $%
l $ are much larger than the coherence length $\xi $. In the limit of low
fraction $\phi \ll 1$ of the SC phase, when the applied Maxwell's approximation is
valid, $l\gg d$. Then our analysis is quantitatively valid at $\xi _{i}\ll
d_{i}$ and gives correct qualitative predictions at $\xi _{i}\lesssim d_{i}$%
. The typical size $d$ of SC islands can be measured for a particular
compound using the STM\cite%
{KresinReview2006,InhBISCCO,InhBISCO2009,InhCaFeAs,InhCaFeAs2018,InhNbN,InhFeSe}
or scanning SQUID microscopy\cite{DiaLaSrCuO,DiaYBCOInh}. In all these
experiments the typical SC domain size $d$ was at least several times larger
than the SC coherence length $\xi $. The smallest SC domain size $d\gtrsim
3nm$ was detected in Bi$_{2}$Sr$_{2}$CaCu$_{2}$O$_{8+\delta }$,\cite%
{InhBISCCO} where the in-plane coherence length $\xi _{ab}\approx 1.6nm$,
thus the ratio $d/\xi \gtrsim 2$. In YBa$_{2}$Cu$_{6}$O$_{y}$ the observed
diamagnetic domain size was much greater,\cite{DiaYBCOInh} $d\sim 1\mu m\gg
\xi _{ab}\sim 2nm$. In NbN the observed SC domains have the size $d\approx
20-50nm\gg \xi \sim 6nm$.\cite{InhNbN} In $\beta $-(BEDT-TTF)$_{2}$I$_{3}$, as we estimated above, the size of SC islands is also $d\sim 1\mu m\gg \xi $. Thus, typically $d/\xi \gg 1$, and
our formulas are applicable. However, we did not find any experimental data
on the domain size in YBa$_{2}$Cu$_{4}$O$_{8}$.

Using torque magnetization measurements, in La$_{2-x}$Sr$_{x}$CuO$_{4}$, Bi$%
_{2}$Sr$_{2-y}$La$_{y}$CuO$_{6}$, Bi$_{2}$Sr$_{2}$CaCu$_{2}$O$_{8+\delta }$
and YBa$_{2}$Cu$_{6}$O$_{y}$ the diamagnetic response as a precursor of
superconductivity was shown to survive at temperatures much higher than the
superconducting transition temperature $T_{c}$.\cite{DiaCuprates} In
particular, in La$_{2-x}$Sr$_{x}$CuO$_{4}$ and Bi$_{2}$Sr$_{2-y}$La$_{y}$CuO$%
_{6}$ the onset temperatures $T_{onset}^{M}$ of this diamagnetic response
exceed more than three times $T_{c}$ in a wide doping intervals and nearly
coincide with the onset temperatures $T_{onset}^{\nu }$ of enhanced Nernst
signal (see Fig. 11 of Ref. [\onlinecite{DiaCuprates}]), presumably corresponding to
the vortex-liquid state. Even in the optimally doped YBa$_{2}$Cu$_{6}$O$_{y}$
with $T_{c}\approx 92$K the diamagnetic response is observed up to $%
T_{onset}^{M}\approx 130$K.\cite{DiaCuprates} According to Ref. [\onlinecite%
{KresinReview2006}], it should coincide with the onset temperature $T^{\ast }$
of pseudogap, which for YBa$_{2}$Cu$_{4}$O$_{8}$ exceeds $200$K. Hence, it
is not very surprising that some traces of superconductivity appear in YBa$%
_{2}$Cu$_{4}$O$_{8}$ at $T\lesssim 200K$, as we see from Fig. \ref%
{fig_phi(T)}. However, we note that the aspect ratio $\gamma\approx 0.14$,
giving the best fit of resistivity curves, coincides within the accuracy of
our model with the ratio of coherence lengths along and perpendicular to
conducting layers. This may indicate that SC fluctuations, probably heterogeneous
and located at the SC islands, may also be partially responsible for the resistivity
drop and nonzero $\phi (T)$ at $T\sim 200$K.

The alternative interpretation of the resistivity decrease in YBa$_{2}$Cu$%
_{4}$O$_{8}$ in the interval $T_{c}<T<T^{\ast }$ is based on the crossover
between coherent metallic at low $T$ and incoherent at high $T$ interlayer
transport.\cite{Hussey:1997} The idea of such a crossover was developed to
explain the nonmonotonic temperature dependence of interlayer conductivity
observed in various layered conductors, including graphite compounds,\cite%
{Graphite} TaS$_{2}$,\cite{TaS2} Sr$_{2}$RuO$_{4}$,\cite{SrRu} organic metals%
\cite{Analytis2006} etc. The most puzzling in this nonmetallic behavior was
that the nonmonotonic temperature dependence of resistivity with a maximum at $%
\sim 100K$ was observed only along the interlayer direction, while the
in-plane conductivity shows metallic behavior. First, this crossover from
coherent to incoherent interlayer transport was believed to happen when the
electron intralayer mean scattering time $\tau $ becomes greater than the
interlayer hopping time $\tau _{z}=\hbar /t_{z}$, so that electrons scatter
many times before tunneling to the adjacent layer. The limit $\tau /\tau
_{z}\ll 1$ received the special term ``weakly incoherent'', but even in
magnetoresistance no considerable changes of behavior have been found at $%
\tau /\tau _{z}\ll 1$.\cite{MosesMcKenzie1999} Later it was realized that
even at $\tau /\tau _{z}\ll 1$ the coherent interlayer transport survives,
and one needs to include the phonon-assisted interlayer tunneling or/and
resonance impurities between the conducting layers into the theoretical
model to explain such behavior.\cite{Abrikosov1999,Maslov,Incoh2009}
In any case, the resistivity decrease at $T_{c}<T<T^{\ast }$
in YBa$_{2}$Cu$_{4}$O$_{8}$ is, probably, mainly due to
the heterogeneous SC onset discussed above rather than due to this coherence-incoherence crossover,
because the analyzed experimental data \cite{Hussey:1997} on the temperature dependence of resistivity in YBa$%
_{2}$Cu$_{4}$O$_{8}$ do not have the resistance maximum, typical for this
coherence-incoherence crossover. Moreover, these data corresponds to the
samples with higher resistance at room temperature, suggesting their strong
spatial inhomogeneity. Of course, both these effects, namely, SC inclusions and the incoherent channels of conductivity may be present and contribute in parallel, leading to the observed decrease of resistivity in YBa$_{2}$Cu$_{4}$O$_{8}$ below $250$K.

The proposed model and analytical results are rather general and can be used for the analysis of experimental data in other strongly anisotropic compounds. Let us briefly summarize the main steps of the comparison of this model with experimental data. First, one chooses a compound where, presumably, superconductivity appears in the form of isolated islands. This is very helpful, but not necessary, if there are STM or other measurements, supporting this heterogeneous SC onset and giving the typical size of SC inclusions. Then one extracts from experimental resistivity data the excess conductivity as a function of temperature along three main axes due to superconducting inclusions. This can be done more easily if there are also experimental data on conductivity in magnetic field or under other conditions, suppressing superconductivity. If the resistivity data without superconducting inclusions are not available, the excess conductivity can be approximately extracted using the extrapolation from higher temperature, where superconductivity is suppressed. These data on excess conductivity along main axes are fitted by the formulas derived above, which gives the temperature dependence of the volume fraction $\phi (T)$ of superconducting inclusions and their aspect ratios $\gamma$ and $\beta$. If, in addition to transport measurements, the diamagnetic response due to superconducting inclusions is measured, it can be used for independent measure of $\phi (T)$. The comparison of $\phi (T)$ from resistivity and susceptibility measurements is helpful to check the consistency and applicability of the proposed model to studied material. It can also be used to estimate the size of superconducting inclusions.

\section{Conclusions}

In this paper we developed a classical model and derived analytical
expressions, given by Eqs. (\ref{eq:rho1})-(\ref{eq:rho3}) and (\ref{A1})-(%
\ref{A3_appr}), for conductivity in a heterogeneous fully anisotropic
conductors with ellipsoid superconducting inclusions. This model and the
analytical results obtained are useful and convenient to analyze
experimental data on the temperature dependence of conductivity anisotropy
in various anisotropic superconductors, where superconductivity onset
happens inhomogeneously in the form of isolated superconducting islands. We
illustrate this by analyzing the experimental data on the temperature dependence of resistivity along
three main axes above the transition temperature $T_c$ in the high-temperature
superconductor YBa$_2$Cu$_4$O$_8$ and in the organic superconductor $\beta$-(BEDT-TTF)$_{2}$I$_{3}$.
In $\beta$-(BEDT-TTF)$_{2}$I$_{3}$ we compared the temperature dependence ofthe  superconductivity fraction extracted from  resistivity and diamagnetic response data, which allows estimating the size of superconducting inclusions as $d\sim 1\mu m$.
We described the comparison between our theory and the experimental data in detail, to make this procedure clear for
applications to other anisotropic superconductors. In spite of its
simplicity, the proposed classical model of anisotropic heterogeneous
superconductor gives a reasonable qualitative and often quantitative
description of the temperature dependence of resistivity and of its
anisotropy above the transition temperature in the compounds with
inhomogeneous superconductivity onset in the form of isolated
superconducting islands.


\begin{acknowledgements}
We thank Yaroslav Gerasimenko and Konstantin Tikhonov for useful discussions.
The paper was partially supported by the Ministry of Education and Science of the Russian Federation in the framework of Increase Competitiveness Program of NUST MISiS and by ``Basis'' Foundation. Sec. V was supported by the Russian Science Foundation (Grant No. 16-42-01100). P.G. acknowledges the program 0033-2018-0001 ``Condensed Matter Physics'' by the FASO of Russia. P.K. acknowledges the support of Arconic Foundation.\end{acknowledgements}


\appendix

\section{Conductivity anisotropy in the standard theory of superconducting
fluctuations}

\label{app:SCFluctuations}

The model in Refs. [\onlinecite{Grigoriev:2017-PRB,GrigorievJETPL2017}] predicts
that if superconductivity in an anisotropic conductors appears in the form
of isolated superconducting islands, it reduces electric resistivity
anisotropically with the maximal effect along the least conducting axis.
This prediction is supported by the experimental data in various compounds.\cite%
{Hussey:1997,DiaBaFeCoAs,Grigoriev:2017-PRB,GrigorievJETPL2017,AnisET2I3,PasquierPRB2010,Gerasimenko2014}
These results cannot be explained by the standard
theory\cite{LarkinVarlamovFluct} of superconducting fluctuations in
homogeneous superconductors, as was argued in Ref. [\onlinecite{GrigorievJETPL2017}].
In this appendix section we briefly repeat these arguments\cite{GrigorievJETPL2017} for completeness and discuss possible extensions of the homogeneous theory of superconducting fluctuations.

According to Chapter 3 of Ref. [\onlinecite{LarkinVarlamovFluct}] within the
time-dependent Ginzburg-Landau equations (i.e., near $T_{c}$), the excess
conductivity due to fluctuations in layered quasi-2D superconductors in the
absence of a magnetic field is given by the expressions
\begin{equation}
\Delta \sigma _{xx}\left( \epsilon ,h=0,\omega =0\right) =\frac{e^{2}}{16s}%
\frac{1}{\sqrt{\epsilon \left( \epsilon +r\right) }},  \label{dsx}
\end{equation}%
\begin{equation}
\Delta \sigma _{zz}\left( \epsilon ,h=0,\omega =0\right) =\frac{e^{2}s}{%
32\xi _{xy}^{2}}\left( \frac{\epsilon +r/2}{\sqrt{\epsilon \left( \epsilon
+r\right) }}-1\right) ,  \label{dsz}
\end{equation}%
where $s$ is interlayer distance, $r=4\xi _{z}^{2}\left( 0\right) /s^{2}$, $%
\xi _{xy}$ and $\xi _{z}$ denote the superconducting coherence length in the
conducting layers and across them, respectively; and $\epsilon \equiv \ln
\left( T/T_{c}\right) \approx \left( T-T_{c}\right) /T_{c}\ll 1$. At $%
\epsilon \ll r$ Eq. (\ref{dsz}) gives
\begin{equation}
\Delta \sigma _{zz}\approx \frac{e^{2}s}{16\xi _{xy}^{2}}\frac{\xi
_{z}^{2}\left( 0\right) /s^{2}}{\sqrt{\epsilon \left( \epsilon +r\right) }}%
=\Delta \sigma _{xx}\frac{\xi _{z}^{2}\left( 0\right) }{\xi _{xy}^{2}},
\label{dsz2}
\end{equation}%
and at $r\ll \epsilon $ from Eq. (\ref{dsz}) we have
\begin{equation}
\Delta \sigma _{zz}=\frac{e^{2}s}{32\xi _{xy}^{2}}\frac{r^{2}}{8\epsilon ^{2}%
}\ll \Delta \sigma _{xx}\frac{s^{2}}{\xi _{xy}^{2}}.  \label{dsz3}
\end{equation}%
In both cases, the excess conductivity across conducting layers $\Delta
\sigma _{zz}$ is much lower (namely, by the parameters $\xi _{z}^{2}/\xi
_{xy}^{2}\ll 1$ or $s^{2}/\xi _{xy}^{2}\ll 1$) than the excess conductivity
along the layers $\Delta \sigma _{xx}$. This small parameter $\xi
_{z}^{2}/\xi _{xy}^{2}\sim v_{z}^{2}/v_{x}^{2}\sim \sigma _{zz}/\sigma _{xx}$%
. Hence, within the Ginzburg-Landau theory, the relative increase of
conductivity due to superconducting fluctuations is isotropic, which cannot
explain the observed\cite%
{Hussey:1997,DiaBaFeCoAs,Grigoriev:2017-PRB,GrigorievJETPL2017,AnisET2I3,PasquierPRB2010,Gerasimenko2014}
temperature dependence of conductivity anisotropy above $T_{c}$.

A stricter microscopic theory of the fluctuation contribution to the
conductivity (see Chapter 7 in [\onlinecite{LarkinVarlamovFluct}] and
references therein) is applicable far away from $T_c$ and includes not only
the Aslamazov-Larkin correction given by Eqs. (\ref{dsx})-(\ref{dsz3}) but
also the Maki-Thompson correction and the correction due to the
renormalization of electron density of states. However, this stricter theory
predicts\cite{ResistancePeakTheory} an increase in the transverse resistance
(not conductivity!) above $T_c$, observed\cite{ResistancePeakExp} in some
cuprates as a resistance peak just above $T_c$. Thus, the much stronger
excess conductivity across the conducting layers cannot be explained within
the existing theory\cite{LarkinVarlamovFluct} of fluctuation conductivity in
spatially homogeneous superconductors.

A possible extension of the existing homogeneous theory\cite{LarkinVarlamovFluct} of SC fluctuations to a
spatially inhomogeneous superconductors, where SC fluctuations appear only in some special spots, may be useful to explain the observed anisotropic correction to resistivity far above $T_c$, where the SC volume fraction according to our model is very small. Such spots of highly probable SC fluctuations somewhat resemble the spots of higher conductivity in our model of SC islands, but instead of steady SC islands with zero resistance one takes islands with reduced resistance due to SC fluctuations. The frequency dependence of conductivity in such a heterogeneous theory of SC fluctuations, probably, differ considerably from that in our model. Such a model of heterogeneous SC fluctuations, being beyond the scope of this paper, may be relevant and useful for superconductors with nonuniform doping concentration, with nonuniform charge- or spin-density wave structure, or with other types of heterogeneity.

\section{Elliptic integrals}

\label{app:Elliptic_integrals}

In this appendix we calculate integrals (\ref{El_int}) and find exact
expressions for coefficients $A_1, A_2, A_3$ as well as their asymptotic
behaviors for different cases.

Let $a_1<a_2<a_3$. Denote $\nu = \arcsin \sqrt{a_3^2-a_1^2}/a_3$ -- --
angular eccentricity; $q = \sqrt{(a_3^2-a_2^2)/(a_3^2-a_1^2)}$; $q^{\prime
}\equiv \sqrt{1-q^2} = \sqrt{(a_2^2-a_1^2)/(a_3^2-a_1^2)}$. Using the table
elliptic integrals (integrals 6, 12, 18 from Sec. 3.133 of Ref. [%
\onlinecite{Gradshteyn-Ryzhik:2007}]) we obtain:
\begin{align}
A_1 =& \frac{a_1 a_2 a_3}{2}\int \limits_0^\infty \frac{dt}{(t+a_1^2)\sqrt{%
(t+a_1^2)(t+a_2^2)(t+a_3^2)}} =  \notag \\
=& \frac{a_1 a_2 a_3}{2} \left( \frac{2}{(a_1^2-a_2^2)\sqrt{a_3^2-a_1^2}}
E(\nu,q) + \frac{2}{a_2^2-a_1^2} \frac{a_2}{a_1 a_3} \right).  \label{A1}
\end{align}

\begin{align}
A_2 &= \frac{a_1 a_2 a_3}{2}\int \limits_0^\infty \frac{dt}{(t+a_2^2)\sqrt{%
(t+a_1^2)(t+a_2^2)(t+a_3^2)}} =  \notag \\
&= \frac{a_1 a_2 a_3}{2} \left( \frac{2\sqrt{a_3^2-a_1^2}}{%
(a_2^2-a_1^2)(a_3^2-a_2^2)} E(\nu,q) \right. -  \notag \\
& \left. \frac{2}{(a_3^2-a_2^2)\sqrt{a_3^2-a_1^2}} F(\nu,q) - \frac{2}{%
a_2^2-a_1^2} \frac{a_1}{a_2 a_3} \right).  \label{A2}
\end{align}

\begin{align}
A_3 &= \frac{a_1 a_2 a_3}{2}\int \limits_0^\infty \frac{dt}{(t+a_3^2)\sqrt{%
(t+a_1^2)(t+a_2^2)(t+a_3^2)}} =  \notag \\
&= \frac{a_1 a_2 a_3}{2} \left( \frac{2}{(a_3^2-a_2^2) \sqrt{a_3^2-a_1^2}}
(F(\nu,q) - E(\nu,q)) \right).  \label{A3}
\end{align}
Here $F(\nu,q)$ and $E(\nu,q)$ are incomplete elliptic integrals of the
first and the second kind respectively with amplitude $\nu$ and the elliptic
modulus $q$ [\onlinecite{Gradshteyn-Ryzhik:2007}]. It can be easily checked
that indeed $A_1 + A_2 + A_3 = 1$.

Let us simplify formulas (\ref{A1})-(\ref{A3}) for two limiting cases: (i) $%
a_3 \gg a_1,a_2$; (ii) $a_2 - a_1 \ll a_3$.

(i) $a_3 \gg a_1,a_2$. In this case $\nu\rightarrow\pi/2$ and $q\rightarrow
1 $. Using the double asymptotic expansions for $F(\nu,q)$ and $E(\nu,q)$ [%
\onlinecite{Kaplan:1948}], we find
%
\begin{equation}
A_{1}\approx \frac{a_{2}}{a_{1}+a_{2}} - \frac{a_1 a_2}{2 a_3^2} \ln \frac{4
a_3/e}{a_1+a_2},  \label{A1_appr}
\end{equation}

\begin{equation}
A_{2}\approx \frac{a_{1}}{a_{1}+a_{2}} - \frac{a_1 a_2}{2 a_3^2} \ln \frac{4
a_3/e}{a_1+a_2},  \label{A2_appr}
\end{equation}

\begin{equation}
A_{3}\approx \frac{a_1 a_2}{ a_3^2} \ln \frac{4 a_3/e}{a_1+a_2}.
\label{A3_appr}
\end{equation}

Substituting here $a_i$ from formula (\ref{eq:shape}), we get
\begin{equation}
A_{1}\approx \frac{\beta}{\sqrt{\mu}+\beta},  \label{A1_appr2}
\end{equation}

\begin{equation}
A_{2} \approx \frac{\sqrt{\mu}}{\sqrt{\mu}+\beta},  \label{A2_appr2}
\end{equation}

\begin{equation}
A_{3}\approx \frac{\beta\eta}{\gamma^2\sqrt{\mu}}\ln \frac{4 \gamma}{e\sqrt{%
\eta} (1+\beta/\sqrt{\mu})}.  \label{A3_appr2}
\end{equation}

(ii) The case $a_2 - a_1 \ll a_3$ (i.e. $a_2-a_1 \rightarrow 0$). In this
case $q\rightarrow 1$ and we use $E(\nu,q) = E(\nu,\sqrt{1-q^{\prime 2}})
\approx \sin\nu + 1/2 (\ln ((1+\sin\nu)/\cos\nu)-\sin\nu) q^{\prime 2}$; $%
F(\nu,q) \approx F(\nu,1) = \ln ((1+\sin\nu)/\cos\nu)$ and obtain

\begin{equation}
A_{1}\approx A_{2}\approx \frac{1}{2}-\frac{a_{1}^{2}a_{3}}{%
2(a_{3}^{2}-a_{1}^{2})^{3/2}}\ln \frac{a_{3}+\sqrt{a_{3}^{2}-a_{1}^{2}}}{%
a_{1}}+\frac{a_{1}^{2}}{2(a_{3}^{2}-a_{1}^{2})}.  \label{A2_appr_ii}
\end{equation}%
%
%
%
%
%
%
%
%
%
\begin{equation}
A_{3}\approx \frac{a_{1}^{2}a_{3}}{(a_{3}^{2}-a_{1}^{2})^{3/2}}\ln \frac{%
a_{3}+\sqrt{a_{3}^{2}-a_{1}^{2}}}{a_{1}}-\frac{a_{1}^{2}}{a_{3}^{2}-a_{1}^{2}%
}  \label{A3_appr_ii}
\end{equation}%
For $a_{1}=a_{2}$ formulas (\ref{A2_appr_ii})-(\ref{A3_appr_ii}) become
exact. In the double limit $(a_{2}-a_{1})\rightarrow 0$ and $%
a_{3}/a_{1}\rightarrow \infty $ we get $A_{3}\sim (a_{1}^{2}/a_{3}^{2})\ln
(2 a_{3}/e a_{1})$, which coincides with (\ref{A3_appr}) when $a_1=a_2$.
Recalling that $a_{1}/a_{3}=\sqrt{\eta _{\ast }}\equiv \sqrt{\eta}/\gamma$
we can recast (\ref{A3_appr_ii}) into
\begin{equation}
A_{3}\approx \frac{\eta _{\ast }}{(1-\eta _{\ast })^{3/2}}\ln \frac{1+\sqrt{%
1-\eta _{\ast }}}{\sqrt{\eta _{\ast }}}-\frac{\eta _{\ast }}{1-\eta _{\ast }}%
,  \label{A3_appr_ii-Grig}
\end{equation}%
which after algebraic manipulations can be transformed into the form of Eq.
(17.30) of Ref. [\onlinecite{Torquato}] or Eq. (5) of Ref. [%
\onlinecite{Grigoriev:2017-PRB}]. For $\eta _{\ast }\rightarrow 0$ the
formula (\ref{A3_appr_ii-Grig}) simplifies to $A_{3}\sim \eta_* \ln
(1/\eta_* )/2$, which is consistent with Eq. (6) of [%
\onlinecite{Grigoriev:2017-PRB}].



\begin{thebibliography}{99}
\bibitem{footnote_open_circles} Data for resistivity for all three axes are available only for filled and
blank circle samples. Of these two, the blank-circle data have been
taken because of its smaller temperature step and less irregularities in the data.
\end{thebibliography}
\end{document}